\documentclass[sigconf]{acmart}

\usepackage{wrapfig}
\PassOptionsToPackage{svgnames}{xcolor}
\usepackage{xcolor}
\usepackage{braket}
\usepackage{soul}
\usepackage{fancyvrb}

\graphicspath{{figs/}{figures/}{pictures/}{images/}{./}} 

\definecolor{myColor}{RGB}{47, 48, 48}

\newcommand*\component[1]{%
  \raisebox{0pt}{\fcolorbox{white}{myColor}{\textsf{\textcolor{white}{#1}}}}%
}
            
\AtBeginDocument{%
  }

\setcopyright{rightsretained}
\copyrightyear{2025}
\acmYear{2025}
\acmDOI{10.1145/3706599.3720085}
\acmConference[CHI EA '25]{Extended Abstracts of the CHI Conference on Human Factors in Computing Systems}{April 26-May 1, 2025}{Yokohama, Japan}
\acmBooktitle{Extended Abstracts of the CHI Conference on Human Factors in Computing Systems (CHI EA '25), April 26-May 1, 2025, Yokohama, Japan}
\acmISBN{979-8-4007-1395-8/2025/04}

\begin{document}

\title{Intuit: \texorpdfstring{Expla\underline{in}} \texorpdfstring{ Quan\underline{tu}m} \texorpdfstring{Comput\underline{i}ng} \texorpdfstring{ Concep\underline{t}s} via AR-based Analogy}

\newcommand{\tbd}[1]{\textcolor{green}{[#1]}}

\newcommand{\lp}[1]{\textcolor{blue}{[linping: #1]}}
\newcommand{\mn}[1]{\textcolor{red}{[manusha: #1]}}
\newcommand{\wy}[1]{\textcolor{brown}{[Yong: #1]}}

\newcommand{\redbox}{\includegraphics[scale=0.04]{icons/Number_of_Qubits.pdf}}
\newcommand{\noofqubits}{\includegraphics[scale=0.04]{icons/Number_of_Qubits.pdf}}
\newcommand{\noofobjects}{\includegraphics[scale=0.04]{icons/Number_of_Objects.pdf}}

\newcommand{\propertycontinuity}{\includegraphics[scale=0.04]{icons/Property_Continuity.pdf}}
\newcommand{\probabilityquantification}{\includegraphics[scale=0.04]{icons/Probability_Quantification.pdf}}

\newcommand{\stateduality}{\includegraphics[scale=0.04]{icons/Output_state_duality.pdf}}
\newcommand{\rotational}{\includegraphics[scale=0.04]{icons/Rotational.pdf}}

\newcommand{\concepttype}{\includegraphics[scale=0.04]{icons/Quantum_Concept_Type.pdf}}
\newcommand{\translational}{\includegraphics[scale=0.04]{icons/Transitional.pdf}}

\newcommand{\toolName}[1]{\textit{Intuit}}
\newcommand{\q}[1]{\textit{``#1''}}

\author{Manusha Karunathilaka}
\authornote{Both authors contributed equally to this research.}
\affiliation{%
  \institution{Singapore Management University}
  \city{Singapore}
  \country{Singapore}
}
\email{gmik.vidana.2023@phdcs.smu.edu.sg}
\orcid{0009-0001-1345-0815}

\author{Shaolun Ruan}
\authornotemark[1]
\affiliation{%
  \institution{Singapore Management University}
  \city{Singapore}
  \country{Singapore}
}
\affiliation{%
  \institution{Monash University}
  \city{Melbourne}
  \country{Australia}
}
\email{haywardryan@foxmail.com}
\orcid{0000-0002-6163-9786}

\author{Lin-Ping Yuan}
\affiliation{%
  \institution{The Hong Kong University of Science and Technology}
  \city{Hong Kong SAR}
  \country{China}}
\email{yuanlp@cse.ust.hk}
\orcid{0000-0001-6268-1583}

\author{Jiannan Li}
\affiliation{%
  \institution{Singapore Management University}
  \city{Singapore}
  \country{Singapore}
}
\email{jiannanli@smu.edu.sg}
\orcid{0000-0001-8409-4910}

\author{Zhiding Liang}
\affiliation{%
 \institution{Rensselaer Polytechnic Institute}
 \city{Troy}
 \state{New York}
 \country{USA}}
\email{liangz9@rpi.edu}
\orcid{0000-0002-7568-0165}

\author{Kavinda Athapaththu}
\affiliation{%
  \institution{Nanyang Technological University}
  \city{Singapore}
  \country{Singapore}}
\email{kavinda.athapaththu@ntu.edu.sg}
\orcid{0000-0002-8641-7768}

\author{Qiang Guan}
\affiliation{%
  \institution{Kent State University}
  \city{Kent}
  \state{Ohio}
  \country{USA}}
\email{qguan@kent.edu}
\orcid{0000-0002-3804-8945}

\author{Yong Wang}
\authornote{Corresponding author.}
\affiliation{%
  \institution{Nanyang Technological University}
  \city{Singapore}
  \country{Singapore}}
\email{yong-wang@ntu.edu.sg}
\orcid{0000-0002-0092-0793}

\renewcommand{\shortauthors}{Karunathilaka et al.}

\begin{abstract}
Quantum computing has
shown great potential to revolutionize traditional computing and can provide an exponential speedup for a wide range of possible applications, attracting various stakeholders. 
However, understanding fundamental quantum computing concepts remains a significant challenge for novices because of their abstract and counterintuitive nature. 
Thus, we propose an analogy-based characterization framework to construct the mental mapping between quantum computing concepts and daily objects,
informed by in-depth expert interviews 
and a literature review,
covering key quantum concepts and characteristics like  number of qubits, output state duality, quantum concept type, and probability quantification. 
Then, we developed an AR-based prototype system, Intuit, using situated analytics to explain quantum concepts through daily objects and phenomena (e.g., rotating coins, paper cutters). 
We 
thoroughly 
evaluated our approach through in-depth user and expert interviews. 
The Results demonstrate the effectiveness and usability of Intuit in helping learners understand abstract concepts in an intuitive and engaging manner.
\end{abstract}


\begin{teaserfigure}
    \setlength{\abovecaptionskip}{-0.005cm} 
  \includegraphics[width=\textwidth]{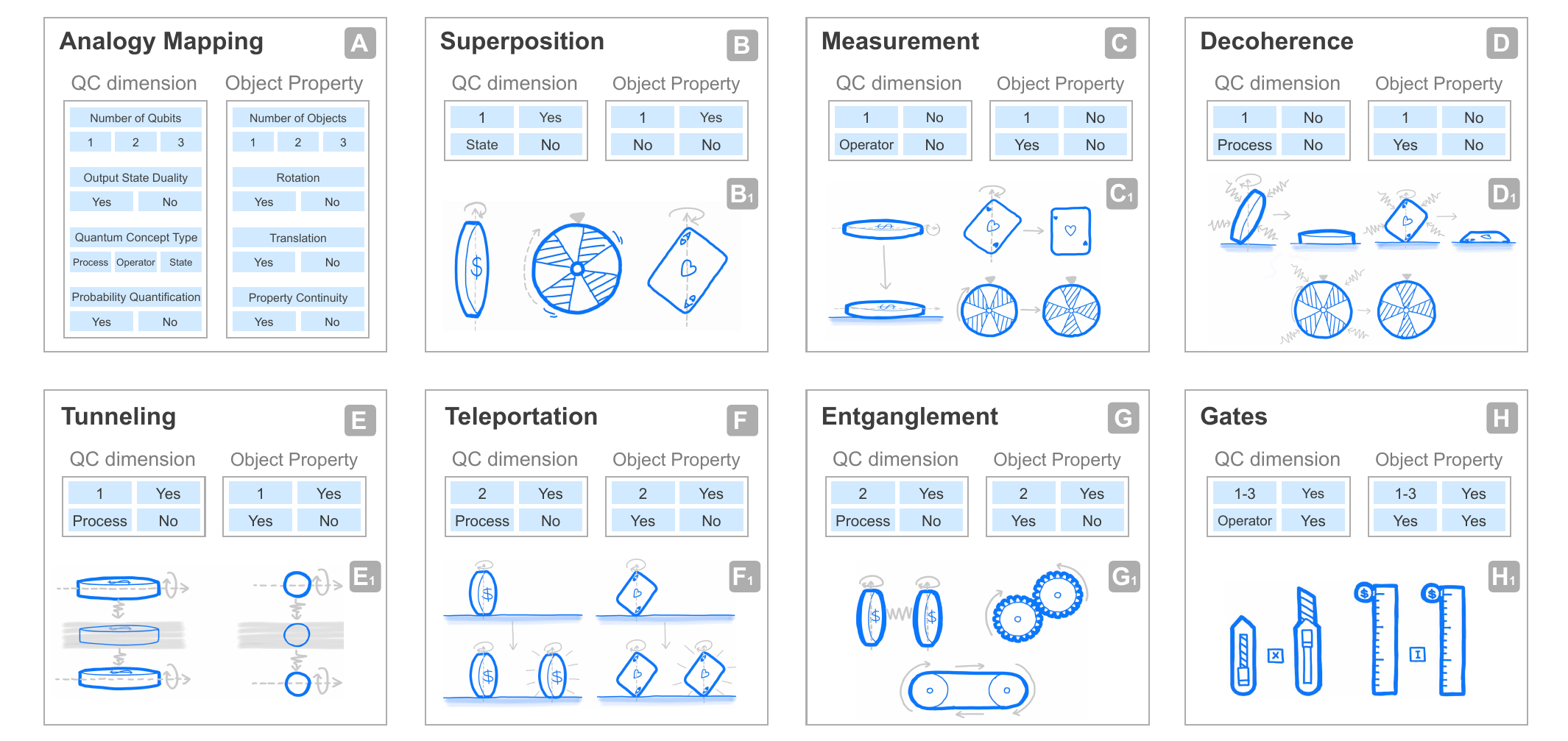}
  \caption{An overview of our analogy-based characterization framework.
    (A) shows the analogy-based mapping between the characterization dimensions of quantum computing (QC) concepts (i.e., number of qubits, output state duality, quantum concept type, and probability quantification) and the corresponding properties of daily objects (i.e., number of daily objects, rotation, translation, and property continuity).
    (B) - (H) showcase the detailed explanation of each quantum computing concept using our analogy-based characterization framework, with several analogy examples provided for each concept. 
    The QC dimensions and object properties shown vertically in (A) correspond to those listed in reading order (left-to-right) in (B) - (H). 
    }
    \label{fig:framework}
\end{teaserfigure}


\maketitle

\section{Introduction}

Quantum computing (QC) is an emerging technology offering exponential speedup over classical computing by leveraging quantum-mechanical principles such as superposition and entanglement~\cite{gill2022quantum,glassner2001quantum,bethel2023quantum,bhat2022quantum}. This computational advantage, known as \textit{quantum advantage}, is expected to solve complex problems in domains like cryptography, materials science, and artificial intelligence. QC has garnered significant interest globally, with countries like the USA, China, and Germany investing heavily in research~\cite{us_national_quantum_initiative,liman2023quantum}. IT leaders such as IBM and Microsoft are advancing QC technologies~\cite{ruan2023quantumeyes}, while academic institutions focus on training future quantum scientists~\cite{fox2020preparing}. Platforms like IBM Quantum and Microsoft Azure Quantum further democratize access to QC tools, fostering broader participation in learning and experimentation.

Learning QC is challenging for novice learners due to its complex and counterintuitive nature~\cite{glassner2001quantum,arida2002quantum,nielsen2010quantum}.
QC principles, such as \textit{superposition}—a system existing in multiple states until measured—and \textit{entanglement}—perfect correlations between distant objects—defy everyday intuition. 
For example, Schrödinger's cat illustrates \textit{superposition}, where the cat is both alive and dead until observed~\cite{schrodinger1935present}. 
Existing educational methods rely heavily on mathematical abstractions, making QC difficult to grasp for beginners~\cite{marshman2015framework}. 
Thus, an intuitive and effective learning approach remains needed.

To fill the above research gap, this work leverages analogy to explain QC concepts, building on prior studies showing that linking new knowledge with familiar objects promotes learning~\cite{gentner1983structure,hofstadter2013surfaces}. 
However, constructing analogies for QC is inherently challenging. 
First, there is no systematic framework or method to identify the suitable familiar objects for explaining QC concepts.
Existing metaphors, like \textit{superposition} explained as “a cat being dead and alive until observed,” are often ad-hoc and lack generalizability. 
Second, QC’s counterintuitive principles often seem like “magic,” making them difficult to visualize. 
To address this, it is crucial to make the “magic” of QC visible and relatable to novice learners.

In this work, we propose a novel analogy-based framework to map QC concepts to daily objects, addressing the challenge of making QC intuitive. 
By collaborating with six domain experts,
we identified key dimensions to characterize fundamental QC concepts, and further propose the corresponding properties required for daily objects to support meaningful mappings.
To enhance engagement, we implemented this framework as an Augmented Reality (AR) application, Intuit, which visualizes QC principles by overlaying quantum behaviors onto familiar objects. 
For instance, Intuit uses a paper cutter, where the slider’s position represents a quantum bit’s probability amplitude, allowing users to grasp QC concepts interactively. 
To the best of our knowledge, this is the first AR-based analogy approach for teaching fundamental QC concepts effectively. 

To evaluate the usefulness and effectiveness of Intuit, we conducted user interviews with 16 participants and expert interviews with 6 domain experts. The results demonstrate that Intuit provides an intuitive and effective way for novice quantum computing learners to understand the fundamental quantum computing concepts. 

The major contributions of this paper can be summarized as follows:

\begin{itemize}
    \item We formulate a systematic analogy-based characterization framework to characterize and explain quantum computing concepts, which is informed by in-depth interviews with domain experts and literature review, and links counterintuitive quantum computing concepts with daily objects and phenomena.

    \item We develop Intuit, a novel AR application to implement the analogy-based characterization framework and visualize abstract quantum computing concepts by combining real-world and virtual objects or phenomena.

    \item We conduct extensive evaluations, including user interviews and expert interviews, to demonstrate the usefulness and effectiveness of Intuit, in explaining quantum computing concepts in an intuitive manner. 
\end{itemize}
\section{Related Work}

\subsection{Quantum Computing Education}

Quantum computing education covers theory, algorithms, hardware, and applications~\cite{aiello2021achieving, uhlig2019generating, salehi2021computer, seegerer2021quantum}. 
Current strategies include gamified approaches, such as quantum tic-tac-toe~\cite{goff2006quantum}, Entanglion~\cite{weisz2018entanglion}, and States and Gates~\cite{carberry2022board}, which simplify quantum principles through gameplay but often face challenges due to the complexity of quantum mechanics and the need for proper guidance.
Similarly, hands-on quantum physics experiments like HQBIT~\cite{casamayou2023hqbit} and quantum circuit coding~\cite{violaris2023physics} provide experiential learning but require specialized equipment and expertise, limiting accessibility. 
While engaging, these methods often lack intuitive designs suitable for novices. 
Intuit addresses these gaps by employing familiar daily object analogies to make quantum computing more approachable and relatable.

\subsection{Visualization for Quantum Computing}

Visualization plays a crucial role in understanding QC concepts due to the fundamental differences between quantum and classical computing~\cite{bethel2023quantum}.
Various visualization techniques and educational tools have been developed to enhance comprehension.
For example, the Bloch Sphere is essential for visualizing single-qubit probability amplitudes and phases~\cite{bloch1946nuclear}, while the Q-sphere~\cite{ibm_quantum_composer} extends this to multi-qubit states. 
Novel visualizations like the satellite plot~\cite{ruan2023venus}, QuantumEyes~\cite{ruan2023quantumeyes}, and VIOLET~\cite{ruan30violet} improve interpretability for quantum circuits and neural networks, and node-link approaches~\cite{miller2021graphstatevis} assist in understanding quantum circuits, including Shor's algorithm~\cite{tao2017shorvis}.
Educational tools such as Google’s Cirq~\cite{google_cirq}, IBM’s Qiskit~\cite{qiskit_api}, and Microsoft’s Quantum Development Kit~\cite{azure_quantum} provide simulated quantum environments, while VR platforms~\cite{zable2020investigating} and AR-based systems~\cite{li2023simulating} offer immersive learning experiences.
However, these methods often fall short of addressing fundamental QC concepts for novices.
Unlike these tools, Intuit focuses on making QC concepts accessible for beginners by using analogies with familiar objects.

\subsection{Situated Analytics and AR for Education}

Situated Analytics (SA) integrates Visual Analytics (VA) and AR, embedding data visualizations into the physical environment to enhance understanding and decision-making~\cite{elsayed2016situated,thomas2018situated,shin2023reality, tong2022exploring}. 
By leveraging SA, this work overlays QC concepts onto real-world objects, allowing learners to interact with abstract phenomena in a tangible context. 
AR-based teaching outperforms traditional methods due to its interactive and cognitive benefits~\cite{hung2017applying,baabdullah2022usage}, showing success in fields like chemistry~\cite{cai2014case}, biology~\cite{arslan2020development}, and physics~\cite{suzuki2020realitysketch}. 
However, AR’s use in QC is underexplored. This work aims to fill that gap by making QC concepts intuitive through AR.

\section{Analogy-based Characterization Framework}

\subsection{Fundamental QC Concept Identification}

To derive fundamental QC concepts,
we first conducted a literature review through textbooks~\cite{mcmahon2007quantum, rieffel2011quantum, hirvensalo2013quantum, bernhardt2019quantum, easttom2021quantum}, online courses (e.g., Coursera, IBM), and research papers~\cite{angara2020quantum, seegerer2021quantum} and distilled terms and concepts. 
Then, we conducted a formative study involving six QC experts, including two professors with over five years of teaching experience and four Ph.D. students specializing in areas like quantum noise mitigation and algorithm optimization. 
The interviews, conducted in person and via video conferencing, gathered insights on key QC concepts for novice learners and explored analogies for abstract concepts. 
Experts identified essential topics, including qubits, superposition, entanglement, measurement, quantum gates, teleportation, decoherence, and tunneling. 
Then, two coauthors conducted a thematic analysis of the study and literature review, identifying a final list of fundamental QC concepts, which was then validated by domain experts.
The complete list of QC concepts along with their definitions is provided in supplementary materials.

\subsection{Characterization Framework}

An analogy, linking the new knowledge with familiar objects or information is helpful for learning the new knowledge~\cite{holyoak1996mental,gick1983schema}, and it involves identifying the similarities between them. 
Thus, we propose a characterization framework that begins by identifying the key dimensions of QC concepts and mapping them to the properties of real-world objects.

\subsubsection{Characterization Dimensions for QC Concepts}
\hfill \break
Quantum computing concepts and real-world objects share characteristics that form analogies for understanding complex quantum computing ideas. 
Over four months of collaboration with experts, including two coauthors, we identified four dimensions to characterize quantum concepts: \textit{\textbf{number of qubits}}, \textit{\textbf{output state duality}}, \textit{\textbf{probability quantification}}, and \textit{\textbf{quantum concept type}}, as shown in Figure~\ref{fig:framework}\component{A}.

\textbf{Number of Qubits}
refers to the number of qubits involved in a concept. 
Single-qubit concepts include superposition, measurement, decoherence, and tunneling, while teleportation and entanglement require two qubits.
Quantum gates include single-qubit gates (e.g., \textit{Pauli-X}) and multi-qubit gates (e.g., \textit{Controlled-NOT}, \textit{Controlled Swap} (3 qubits)).

\textbf{Output State Duality}
describes whether the concept’s output state is \textit{dual} (e.g., $\ket{0}$ and $\ket{1}$) or \textit{non-dual} (e.g., $\ket{0}$ or $\ket{1}$). 
For example, superposition, entanglement, tunneling, teleportation, and gates are \textit{dual} due to their probabilistic states, while measurement and decoherence produce \textit{non-dual} outputs by collapsing to a single state.

\textbf{Quantum Concept Type}
includes three categories: \textbf{\textit{state}}, \textbf{\textit{process}}, and \textbf{\textit{operator}}. \textit{State} refers to static conditions, unchanging over
time (e.g., superposition). 
\textit{Process} involves transformations from one state to another (e.g., decoherence, tunneling, teleportation, entanglement). 
\textit{Operator} performs actions on states, such as measurement and quantum gates.

\textbf{Probability Quantification}
distinguishes between concepts that explicitly involve quantum state probabilities (\textit{Probabilistic}, e.g., quantum gates with probability amplitudes) and those that focus on state behavior without explicit probability calculus (\textit{Non-Probabilistic}, e.g., superposition, measurement, decoherence, teleportation, tunneling, entanglement).

\subsubsection{Mapping to Properties of Daily Objects}
\hfill \break
The analogy-based framework maps four quantum dimensions to four daily object properties: \textbf{\textit{number of objects}}, \textbf{\textit{rotation}}, \textbf{\textit{translation}}, and \textit{\textbf{property continuity}} (Figure~\ref{fig:framework}\component{A}).

\textbf{Number of Objects}
corresponds to the \textit{Number of Qubits}. 
Single-object analogies represent single-qubit concepts (e.g., superposition, measurement), while multi-object analogies represent concepts requiring multiple qubits (e.g., entanglement, teleportation). Quantum gates are represented by the number of objects corresponding to their input qubits.

\textbf{Rotation}
corresponds to the \textit{Output State Duality}. 
Rotating objects symbolize \textit{dual} states (e.g., superposition, tunneling), while stationary objects represent non-dual states (e.g., measurement, decoherence), illustrating quantum uncertainty.

\textbf{Translation}
determines whether an object’s state changes (e.g., stop rotation), aligning with the \textit{Quantum Concept Type}. 
\textit{State} concepts (e.g., superposition) require no change (``No'' in our framework), while \textit{process} and \textit{operator} concepts (e.g., measurement, decoherence, gates) indicate translation, reflecting a state change (``Yes'' in our framework).

\textbf{Property Continuity}
corresponds to the \textit{Probability Quantification}, where the probabilistic nature of concepts necessitates a continuous approach to represent. 
For example, \textit{probabilistic} concepts like gates require objects to represent continuous probabilities, while \textit{non-probabilistic} concepts (e.g., superposition, measurement) can use discrete objects.

\section{Prototype System}

Intuit uses a client-server architecture with two components: \textit{IntuitVision} (server-side) and \textit{IntuitSense} (client-side). 
\textit{IntuitVision}, a Python server, handles object detection using computer vision, while \textit{IntuitSense}, a Unity application, synchronizes user inputs with the virtual environment.
Working through multiple iterations of implementation and collaborating with QC and AR/VR experts, Intuit provides text definitions, math equations, auditory support, daily object interactions, and gesture-based interaction capabilities. 

\subsection{Interaction}

Our prototype incorporates gesture recognition and object tracking to enable intuitive interactions. 
Using a YOLOv8-trained model~\footnote{\url{https://docs.ultralytics.com/}}, we track the cubes and the paper cutter for quantum gate lessons, supporting real-time manipulation. 
Bare-hand tracking allows natural interaction with two gestures: a fist to execute actions (e.g., placing a coin in superposition) and a thumbs-up to return to the main menu.

\subsection{Interactive Modules and Familiar Objects}
\label{sect-dailyobjects}

We developed seven interactive modules with nine lessons covering fundamental quantum concepts, allocating three lessons to Identity, Pauli-X, and Hadamard gates.
Lessons use tangible objects like coins, paper cutters (without the blade), and color-coded cubes (e.g., red cube for Pauli-X gates) to make abstract concepts more engaging and accessible.

\begin{figure*}[h]
  \centering 
  \setlength{\belowcaptionskip}{-0.8em}
  \includegraphics[width=0.8\textwidth
  ]{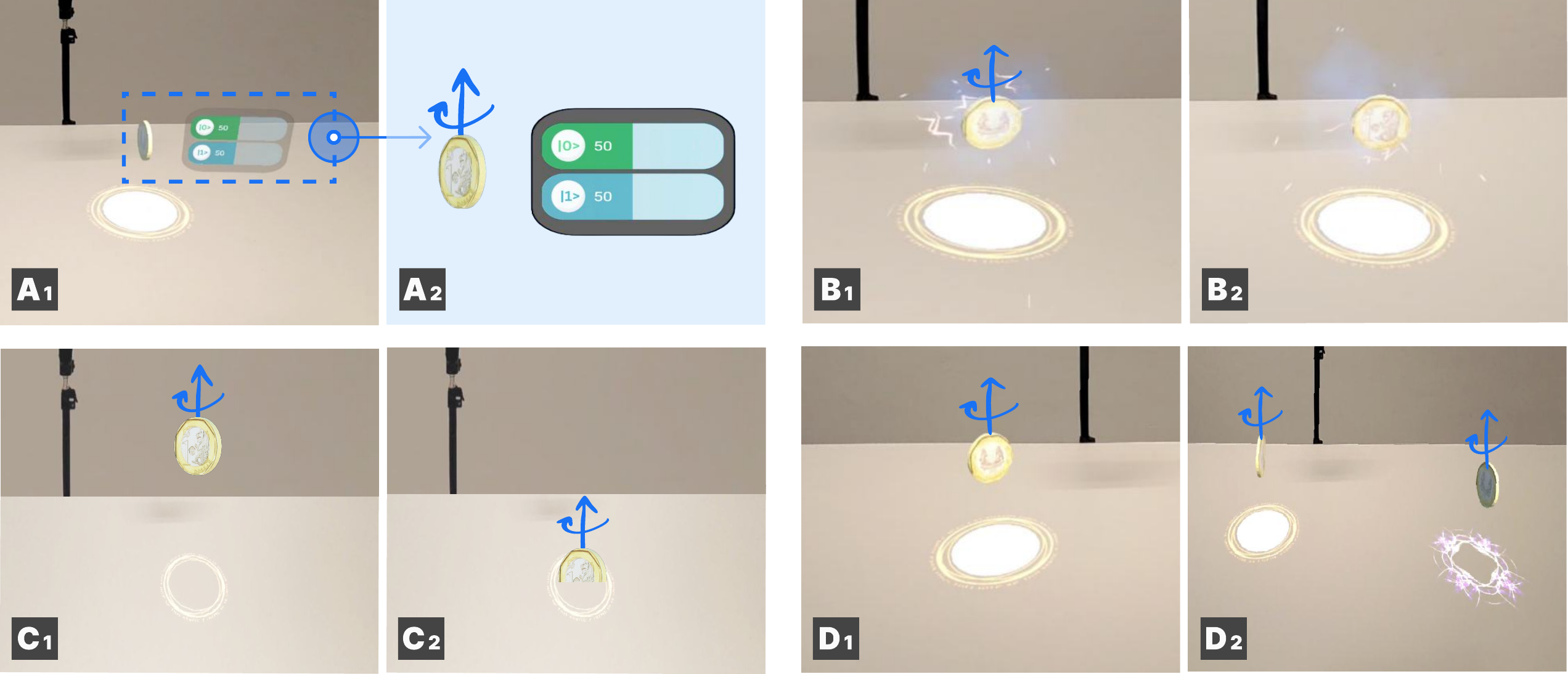}
  \caption{
  \textbf{Superposition} (A): 
  (A\textsubscript{1}) real system indicates the coin is rotating around itself;
  (A\textsubscript{2}) the rotating coin and related probability in probability panel as 50\% for both $\ket{0}$ and $\ket{1}$.
  \textbf{Decoherence} (B):
  (B\textsubscript{1}) the system shows an animation of a rotating coin interacting with the environment;
  (B\textsubscript{2}) over time, the rotation stopped and the coin began to behave like a classical bit. 
  \textbf{Tunneling} (C): a rotating coin passes through a table indicating passing through a barrier (C\textsubscript{1} - C\textsubscript{2}). 
  \textbf{Teleportation} (D): 
  (D\textsubscript{1}) a rotating coin indicates that the coin is in a superposition; 
  (D\textsubscript{2}) an animation illustrates the state transfer to another coin at a different location.
  }
  \label{fig:examples}
\end{figure*}

To complement real objects, we include virtual counterparts to simplify abstract concepts. 
For example, in the superposition lesson (Figure~\ref{fig:examples}\component{A}), a real one-dollar coin is paired with its virtual equivalent. 
A probability panel (Figure~\ref{fig:examples}\component{A\textsubscript{2}}) shows quantum state probabilities, such as 50\% for both |0⟩ and |1⟩ in superposition. 
Similarly, the paper cutter serves as an input mechanism for gate simulations.

\textbf{Superposition} (Figure~\ref{fig:framework}\component{B}) is represented by a rotating object, reflecting dual states. 
Examples include a spinning coin, playing card, or spinner wheel with two colors (Figure~\ref{fig:framework}\component{B\textsubscript{1}}). 
While rotating, these objects simultaneously embody multiple states, illustrating the essence of superposition and offering a tangible way to understand this concept.
In the lesson (Figure~\ref{fig:examples}\component{A}), a rotating coin illustrates quantum state ambiguity, enabling users to simulate superposition.

\begin{figure*}[h]
  \centering 
   \setlength{\abovecaptionskip}{0.1cm} 
  \setlength{\belowcaptionskip}{-0.9em}
  \includegraphics[width=1\textwidth
  ]{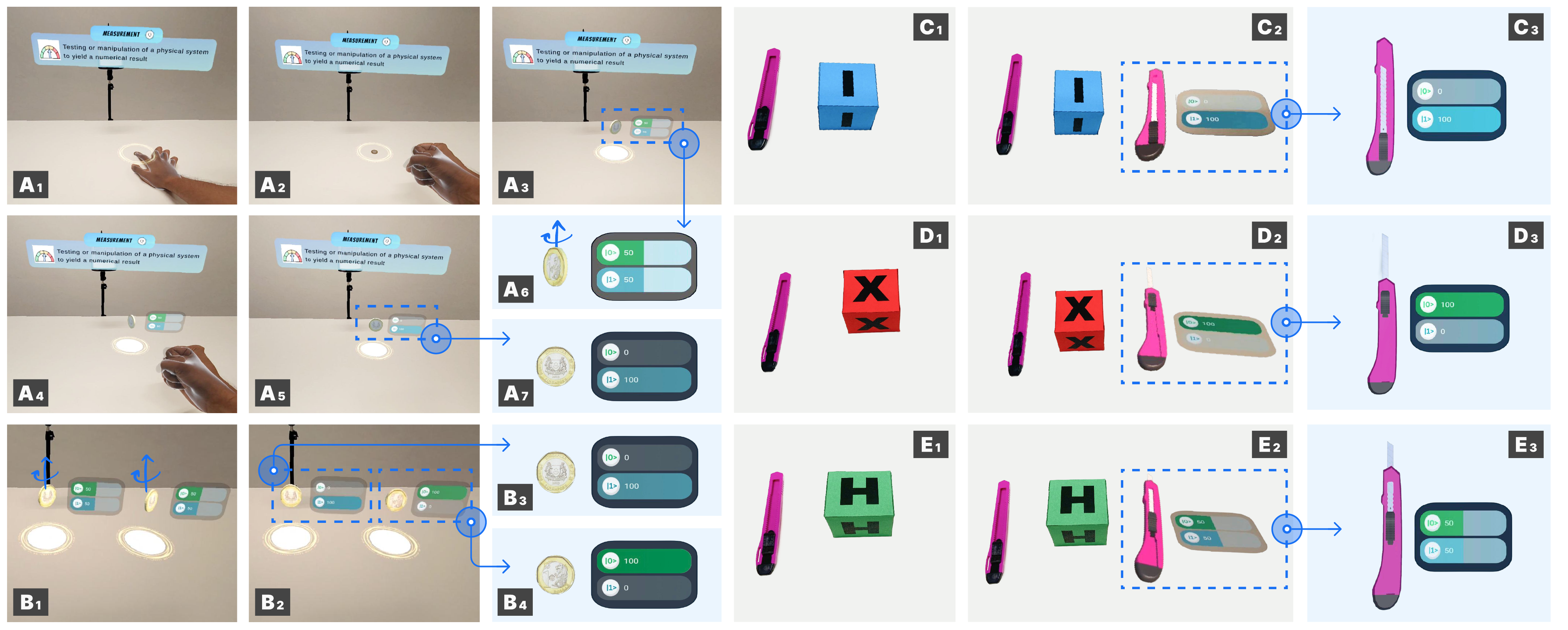 }
  \caption{
  \textbf{Measurement} (A): 
    (A\textsubscript{1}) user places the coin inside a ``magic circle'';
    (A\textsubscript{2}) user makes a fist gesture to trigger superposition;
    (A\textsubscript{3}) the physical coin "disappears" and the system explains that the coin is in superposition using a virtual rotating coin; 
    (A\textsubscript{6}) enlarged probability panel shows 50\% for $\ket{0}$ and $\ket{1}$;
    (A\textsubscript{4}) user performs another fist gesture to measure the state; 
    (A\textsubscript{5}) coin becomes either head or tail (e.g., tail mapped to $\ket{1}$), collapsing the superposition; 
    (A\textsubscript{7}) enlarged probability panel shows the measured state (e.g., tail).
  \textbf{Entanglement} (B):
  (B\textsubscript{1}) two coins are rotating, indicating both coins are in superposition; 
  (B\textsubscript{2}) user measures the left coin, stopping its rotation and causing it to become tail ($\ket{1}$), 
  while the right coin becomes head ($\ket{0}$) at the same time; 
  (B\textsubscript{3}) left coin: tail, 100\% for $\ket{1}$; 
  (B\textsubscript{4}) right coin: head, 100\% for $\ket{0}$.
  \textbf{Gates} ((C) - (E)): 
  gates section includes \textbf{Identity gate} (C), \textbf{Pauli-X gate} (D), \textbf{Hadamard gate} (E).
  (C\textsubscript{1}, D\textsubscript{1}, E\textsubscript{1}) represents the user placing a paper cutter and the color-coded cube for each gate. Input probabilities are  0\% for $\ket{0}$ and 100\% for $\ket{1}$ based on the paper cutter's slider position.
  (C\textsubscript{2}, D\textsubscript{2}, E\textsubscript{2}) shows virtual counterpart of the paper cutter showing the output probabilities.
  (C\textsubscript{3}): the output matches the input, showing that the Identity gate makes no change.
  (D\textsubscript{3}): the output is inverted, showing that the Pauli-X gate flips the state.
  (E\textsubscript{3}): 50\% for both $\ket{0}$ and $\ket{1}$, showing that the Hadamard gate creates a superposition.
  }
  \label{fig:gates_ex}  
\end{figure*}

\textbf{Measurement} (Figure~\ref{fig:framework}\component{C}) is represented by a singular qubit collapsing into a non-dual state, defined as an operator. 
Analogies include halting a rotating coin, card or spinner, transitioning from ambiguity to a single state (Figure~\ref{fig:framework}\component{C\textsubscript{1}}).
The lesson extends the superposition concept by collapsing the coin’s rotation to either head or tail, highlighting the inherent unpredictability of measurement.
Figure~\ref{fig:gates_ex}\component{A} illustrates a usage scenario of the system involving measurement.

\textbf{Decoherence} (Figure~\ref{fig:framework}\component{D}) reflects a singular qubit transitioning to a non-dual state due to environmental interactions, defined as a process. 
Examples include a rotating coin, card or spinner slowing and halting over time as potential energy dissipates due to environmental effects (Figure~\ref{fig:framework}\component{D\textsubscript{1}}).
The lesson in Figure~\ref{fig:examples}\component{B} uses a coin analogy to illustrate how environmental effects slow a spinning coin, leading it to adopt classical bit behavior, demonstrating decoherence.

\textbf{Tunneling} (Figure~\ref{fig:framework}\component{E}) is a process involving a singular qubit in a dual state passing through a barrier. 
Analogies include a virtual object (e.g., coin, playing card) moving through a barrier (e.g., table, wall) (Figure~\ref{fig:framework}\component{E\textsubscript{1}}).
The lesson shown in Figure~\ref{fig:examples}\component{C}, uses a virtual demonstration with a rotating coin that seemingly passing through a table to illustrate quantum tunneling, a phenomenon where quantum particles overcome barriers that would be impassable classically.

\textbf{Teleportation} (Figure~\ref{fig:framework}\component{F}) is a process involving two qubits in a dual state, representing the transfer of information between locations. 
An example is a rotating coin transferring its rotation angle to another coin in a different location (Figure~\ref{fig:framework}\component{F\textsubscript{1}}).
The lesson shown in Figure~\ref{fig:examples}\component{D}, utilizes a coin analogy, depicting through animation how a rotating coin transfers quantum information, resulting in a coin at a different location with transferred quantum information.

\textbf{Entanglement} (Figure~\ref{fig:framework}\component{G}) refers to a process with two qubits in a dual state, representing correlation between objects. 
Examples include two rotating coins where measuring one determines the other, or connected gears rotating in opposite directions (Figure~\ref{fig:framework}\component{G\textsubscript{1}}).
The lesson in Figure~\ref{fig:gates_ex}\component{B} uses two coins to illustrate entanglement. 
Both coins start in a superposition, represented by the rotation. 
Measuring the left coin collapses its state to heads or tails, while the right coin simultaneously collapses to the opposite state, illustrating one possible scenario of entanglement.

\textbf{Gates} (Figure~\ref{fig:framework}\component{H}) involve 1--3 qubits and reflect dual states as operators tied to probability amplitudes. 
Analogies include a paper cutter or a combination of a ruler and coin, representing input qubits and their probabilistic operations (Figure~\ref{fig:framework}\component{H\textsubscript{1}}).
Gates module includes lessons on Identity (Figure~\ref{fig:gates_ex}\component{C}), Pauli-X (Figure~\ref{fig:gates_ex}\component{D}), and Hadamard (Figure~\ref{fig:gates_ex}\component{E}) gates, using the paper cutter analogy with color-coded cubes. 
The slider's position maps probabilities for $\ket{0}$ and $\ket{1}$, where the bottom corresponds to 0\% probability for $\ket{0}$, and the top to 100\% for $\ket{0}$. 
As the slider moves, probabilities adjust, always summing to 100\%.
The color-coded cubes represent gate types: blue with "I" for Identity, red with "X" for Pauli-X, and green with "H" for Hadamard gates. 
Users adjust the slider to observe changes in the virtual paper cutter and probability panel. 
Furthermore, mathematical expressions are included to explain matrix multiplication in gates.

\section{Evaluation}

\subsection{User Interviews}

\begin{figure*}[h]
    \centering
 \setlength{\abovecaptionskip}{0.1cm} 
  \setlength{\belowcaptionskip}{-0.8em}
    \includegraphics[width=\textwidth]{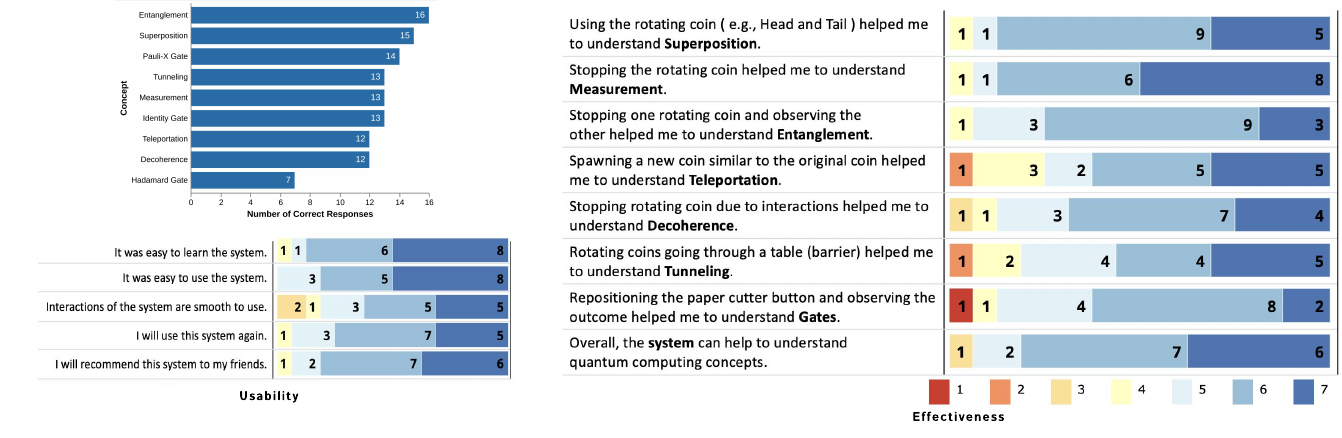}
    \caption{
  Study results: (top-left) number of correct responses over each concept; 
  (right) the effectiveness ratings of the visualizations for each concept. 1=strongly disagree and 7=strongly agree;
  (bottom-left) the usability ratings of the whole system.
  }
    \label{fig:results_n_gates}
\end{figure*}

We conducted user interviews to evaluate the effectiveness of individual visualizations for each concept and the usability of the overall experiences of Intuit.
Our goal is to gather rich qualitative feedback and to determine whether Intuit on its own could effectively aid in understanding quantum concepts, rather than how it performs relative to another method.
The study was approved by our university’s IRB.

\subsubsection{Study Setup and Design}
\hfill \break
\textbf{Participants.}
We recruited 16 participants (P1-P16; 12 males; age range: 21-33) at our university. 
Among the participants, 15 of them were Ph.D. students and the other was a non-Ph.D. student.
Their self-reported familiarity with quantum computing was 2.7 (\textit{i.e.}, they heard of it but do not really know the details) on average on a 7-point Likert scale, where 
1 indicates the least familiarity (\textit{i.e.}, never heard of it) and 7 indicates a high level of knowledge. 
Since the participants had little prior knowledge, we decided to omit pre-test in our study.
All participants had prior AR experience, allowing them to focus on learning QC concepts instead of struggling with an unfamiliar technology. 
We acknowledge this limits insights into AR-naive users, a direction for future work.

\textbf{Procedures.}
Each participant individually underwent a study session starting with an introduction to study goals and procedures.
They were trained on qubit property mappings using objects like coins and paper cutters and learned to interact with the system via hand gestures. 
Participants explored the nine lessons through interactive visualizations over 20 minutes, followed by a quiz with nine questions curated by a quantum computing expert. 
They rated visualization effectiveness and system usability on a 7-point Likert scale and provided feedback on improvements. 
Each session lasted 45--60 minutes, and the participants received SGD 15.
The study was conducted in a lab setting using a Quest 3 device along with an external camera to capture video feedback for object tracking, as Meta's privacy policies limit access to the inbuilt camera's data.

\subsubsection{Results}
\hfill \break
\textbf{Number of correct responses.}
Figure~\ref{fig:results_n_gates} shows the correct answers from 16 participants for each concept. All participants answered entanglement correctly, but the Hadamard gate had the lowest performance, with only 7 correct responses.
Two main reasons for the mistakes emerged. First, the participants found it is hard to recall and differentiate the three gates (Identity, Pauli-X, and Hadamard), with several noting that the Hadamard Gate was especially challenging (e.g., P4, P12). Second, terminology confusion led to errors, with many participants confusing terms like tunneling and teleportation (e.g., P4).

\textbf{Effectiveness.} 
Figure~\ref{fig:results_n_gates} shows the effectiveness ratings of visualizations. Superposition, measurement, and entanglement received ratings above 4/7 from all participants, indicating the rotating coin analogy was helpful. Other concepts also received mostly positive ratings, though P15 criticized the gates, citing delayed response in the Hadamard gate. 

\textbf{Usability.} 
Figure~\ref{fig:results_n_gates} shows usability ratings, with all dimensions rated positively except for interactions, where two participants responded somewhat disagreed on their smoothness, complaining about the recognition accuracy of hand gestures rather than the system design. 
P14 commented that precise gestures were sometimes difficult to perform reliably.

\textbf{Advantages.} 
Participants appreciated the system as an interactive learning tool. 
P4 found it effective for understanding quantum basics, while P1 noted that using daily objects like coins made the concepts relatable and memorable. 
P7 highlighted AR’s value in visualizing complex concepts, and P10 praised the clarity of animations.

\textbf{Suggestions.} 
Participants suggested more control over animations, with P13 requesting adjustable speeds for better observation. They also wanted diverse examples for concepts (e.g., P4) and better integration of visualizations with mathematical explanations, as P11 recommended starting with non-mathematical examples followed by matrix derivations.

\subsection{Expert Interviews}

We conducted remote Zoom interviews with 6 domain experts (E1-E6), with varying experience in QC (2-7 years), who did not participate in the formative study, to evaluate the usefulness of Intuit. 
Notably, E1, E2, E5, and E6 had prior teaching experience in quantum computing.
The experts were not compensated for their participation.
Since the participants were located remotely, the interviews were conducted online via Zoom, and they couldn't interact with Intuit directly.
Instead, we presented demo videos on the lessons and collected feedback by asking open-ended questions, with each session lasting 30–45 minutes. 
The feedback was largely positive, particularly regarding the use of analogies and AR for teaching QC. 
All the experts highlighted the importance of fundamental QC concepts and the difficulty of teaching quantum concepts to others, while appreciating the analogy-based explanations (e.g., E1-3).
E3 recommended enhancing the Hadamard gate lesson with an analogy involving equal but distinct parts, showing two virtual paper cutters as output. 
E5 suggested adding more details to lessons, such as tunneling and teleportation, including concepts like no-cloning theorem to clarify the impossibility of replicating an unknown quantum state.
While acknowledging the current fundamental concepts, experts recommended extending our characterization framework to other concepts, which we reserved for future work.
\section{Discussion}

Our study demonstrates that using analogies with everyday objects like rotating coins effectively simplifies QC concepts such as superposition and measurement, making abstract concepts more relatable. 
However, the Hadamard gate remained challenging, indicating the need for clearer visual distinctions between gates. 
Experts emphasized the potential of using AR with real-world analogies to teach quantum computing, highlighting how this approach could make complex concepts more relatable and engaging, paving the way for future exploration.
Gesture-based interactions were appreciated, though recognition accuracy posed usability challenges, underscoring the importance of balancing immersive interactions with reliability. 
Expanding the framework to cover more advanced QC concepts, such as quantum circuits and quantum neural networks, and incorporating adaptive learning features could improve learning outcomes and enhance the system's generalizability. 

\section{Conclusion}

This paper introduces a systematic analogy-based framework to explain quantum computing concepts, developed in collaboration with domain experts. 
We implemented this framework as Intuit, an AR application that combines real and virtual objects to visualize abstract concepts such as superposition, measurement, decoherence, tunneling, teleportation, entanglement, and gates. 
Evaluations, including user and expert interviews, demonstrate Intuit’s effectiveness and usefulness in facilitating intuitive understanding of quantum computing concepts. 
We hope this work inspires future research on using immersive techniques to  facilitate the learning and education of quantum computing.

\bibliographystyle{ACM-Reference-Format}
\bibliography{sample-base}

\end{document}